\documentclass[materials,article,accept,moreauthors,pdftex]{mdpi} 

\firstpage{1} 
\pubvolume{1}
\issuenum{1}
\articlenumber{0}
\pubyear{2021}
\copyrightyear{2021}
\externaleditor{Academic Editor: {Shin-Hyun Kim} 
} 
\datereceived{23 March 2021} 
\dateaccepted{19 April 2021} 
\datepublished{} 
\hreflink{https://doi.org/} 

\Title{Off-Lattice Monte-Carlo Approach for Studying Nucleation and Evaporation Phenomena at the Molecular Scale}
\TitleCitation{Off-Lattice Monte-Carlo Approach for Studying Nucleation and Evaporation Phenomena at the Molecular Scale}

\Author{ 
Panagiotis E. Theodorakis $^{1,}$*\orcidA{}, Yongjie Wang $^{1}$\orcidB{}, Aiqiang Chen $^{2}$ and Bin Liu $^{2,}$*}

\AuthorNames{Panagiotis E. Theodorakis, Yongjie Wang, Aiqiang Chen and Bin Liu}
\AuthorCitation{Theodorakis, P.E.; Wang, Y.; Chen, A.; Liu, B.}
\address{%
$^{1}$ \quad Institute of Physics, Polish Academy of Sciences, Al. Lotnik\'ow 32/46, 02-668 Warsaw, Poland; {wang@ifpan.edu.pl} (Y.W.) 
	 \\
$^{2}$ \quad Tianjin Key Laboratory of Refrigeration Technology, Tianjin University of Commerce, Tianjin 300134,  China; {chenaiqiang@tjcu.edu.cn} (A.C.)	}

\corres{Correspondence: panos@ifpan.edu.pl (P.E.T.); lbtjcu@tjcu.edu.cn (B.L.)}

\abstract{Droplet nucleation and evaporation are ubiquitous in nature and many technological
applications, such as phase-change cooling and boiling heat transfer. So far, the description of
these phenomena at the molecular scale has posed challenges for modelling with most of the models
being implemented on a lattice. Here, we propose an {off-lattice}
 Monte-Carlo approach
combined with a grid that can be used for the investigation of droplet formation and evaporation.
We provide the details of the model, its implementation as  Python code, and results illustrating
its dependence on various parameters. The method can be easily extended for any force-field
({e.g.,} coarse-grained, all-atom models, and external fields, such as gravity and electric
field). Thus, we anticipate that the proposed model will offer opportunities for a wide range of
studies in various research areas involving droplet formation and evaporation and will also form
the basis for further method developments for the molecular modelling of such phenomena.
}


\begin{document}


\section{Introduction}

Nucleation and evaporation of liquid droplets~\cite{MacDowell2004,Holyst2013,Saenz2017,Sefiane2011b,Brutin2018} 
in contact with solid substrates are ubiquitous in nature ({e.g.,} condensation and
evaporation of rain droplets) and many technological applications, such as phase-change cooling
~\cite{Sefiane2011} and boiling heat transfer~\cite{Kim2010}. These phenomena are complex and 
they can be influenced by different factors, such as thermodynamic conditions~\cite{Chen2015},
substrate properties~\cite{Stauber2015}, and~impurities~\cite{Park2012}. Even when one considers
the simplest case of droplets without impurities onto unstructured and smooth substrates under
constant thermodynamic conditions, the~understanding of these processes continues to pose
challenges. The~origin of these phenomena lies in the interactions
among the molecules of the system, which is difficult to capture with experimental or
theoretical methods and continuum simulations. Certain challenges also exist in the 
case of molecular modeling, as~these phenomena are non-equilibrium processes that require the
exchange of molecules between the liquid drop and the surrounding environment. To~this end,
various molecular models have been proposed over the last years, which have aimed at overcoming 
those challenges and have also led to relevant investigations of these fascinating phenomena~\cite{Crivoi2014,Zhang2016,Andersen2019,Zhang2015,Areshi2019,Nie2017,Cachile1999,Rabani2003,Sztrum2005,Pauliac2008,Vancea2008,Stannard2008,Filipponi2016}. 
For example, molecular-level studies have focused on evaporating droplets with
nanoparticles. In~this case, as~the liquid evaporates, various patterns form, which, in turn, 
can be compared with experimental observations~\cite{Crivoi2014,Zhang2016,Rabani2003, Liu2020}. 
Despite these studies, a~more accurate description of nucleation and evaporation phenomena requires the development of new simulation frameworks, even for simple systems, for~example,
single-component systems without the presence of external~fields.

Fillipponi and Giammatteo have investigated the classical nucleation process by using kinetic
Monte-Carlo simulation (KMC) \cite{Filipponi2016}. Their approach was applied for a wide range
of temperatures providing descriptions in line with the classical nucleation theory, in~particular with respect to parameters describing the average population distribution of the
nuclei size. However, this approach is stochastic in nature, requiring an approximation to the
exact dynamics by generating a set of random integer numbers from Poisson distributions, which
is also computationally demanding. A~similar approach has been employed to study the
nucleation-growth processes of transition metal dichalcogenides~\cite{Nie2017}. Reducing
the volume of a system or using schemes based on the grand canonical ensemble one could simulate
the droplet formation~\cite{MacDowell2004}. In~the latter approach, however, the~formation is spontaneous,
which limits a detailed investigation of the droplet nucleation~mechanisms.

Various molecular models have been proposed to investigate droplet evaporation processes.
Zhang~et~al. \cite{Zhang2015} have employed molecular dynamics simulations of 
all-atom force-fields to investigate the wetting and evaporation of salt-water nanodroplets 
on platinum surfaces focusing on the patterns formed as a result of evaporation. Although~this
method offers an accurate atomistic description of the system, it is computationally demanding
and requires particular care for carrying out the simulations and their subsequent analysis. 
In contrast, other molecular models, such as those based on Monte-Carlo (MC) techniques, can
overcome such limitations~\cite{Zhang2016,Rabani2003,Sztrum2005,Areshi2019}. In~particular, 
MC models offer flexibility in choosing the various moves that would transform the system from one
state to the next (including attempts to remove or add particles) and confirming the acceptance
of such moves by using the Metropolis criterion. For~example, based on the latter approach,
Rabani~et~al. have created a two-dimensional (2D) \cite{Rabani2003} and a
three-dimensional (3D) \cite{Sztrum2005} model for a liquid droplet laden with nanoparticles,
which has later been applied in further applications, such as the study of instabilities in
dewetting nanofluids~\cite{Pauliac2008,Vancea2008} and patterns obtained from drying colloidal
nanoparticle solutions~\cite{Stannard2008}. A~more recent version of the 2D version of the
Rabani model~\cite{Rabani2003} considers a chemical potential that depends on time and the
radius of the droplet. In~this study, Zhang~et~al. have found different drying patterns 
that are in good agreement with experimental results~\cite{Zhang2016}. Finally, a~similar MC
approach on the lattice and the link to hydrodynamics has been discussed in a recent study by
Areshi~et~al. \cite{Areshi2019}. A~common feature of all the above MC studies is the
use of lattice models, which considerably simplifies the implementation of the applied method
and overcomes various difficulties in dealing with the exchange of particles at the 
liquid--vapor interface between the droplet and the surrounding vapor. 
However, if~we would like to better capture the dynamic behavior of the nucleation
and the evaporation of a droplet, an~off-lattice approach would be desirable along with a more
natural representation of the system (flexibility in choosing the force-field), 
especially close to the droplet surface where these phenomena~manifest.

Here, we address these issues by proposing an off-lattice MC approach for studying droplet nucleation
and evaporation phenomena. The~approach is based on a standard off-lattice MC scheme in the
canonical ensemble for the bulk of the droplet, 
which is additionally equipped with the ability of removing and adding
particles at the liquid--vapor (LV) interface by using a suitable grid and the chemical 
potential. Here, the~implementation of the model is illustrated in a simple system of Lennard-Jones (LJ) particles,
but it can easily be extended for systems that include nanoparticles or other molecules.
Moreover, the~developed method can be used with any available forcefield, be it all-atom or
coarse-grained. Hence, we anticipate that our approach will form the basis for further
conceptual developments in this area. In~the following, we discuss the method details and
provide a parametric study of the proposed model. An~implementation of the model as
Python code is available as Supplementary~Information.

\section{Simulation Model and~Method}
Our system consists of an implicit substrate and a droplet of coarse-grained beads that
interact by means of the LJ 12--6 potential:
\begin{equation}\label{eq:LJpotential126}
U^{\rm 12-6}(r) = 4\varepsilon_{\rm d} \left[  \left(\frac{\sigma_{\rm  d}}{r}
\right)^{12} - \left(\frac{\sigma_{\rm d}}{r}  \right)^{6}    \right].
\end{equation}

\noindent
Here, only interactions between beads at distances, $r$, smaller than a cutoff distance are
considered. This cutoff is $r_{\rm c}=2.5\sigma$, where $\sigma$ is the unit of length. The~
LJ potential is also shifted at the cutoff. As~a result the energy $U^{\rm 12-6}(r_{\rm c})=0$.
The parameter $\varepsilon_{\rm d}$ tunes the strength of the LJ interaction between the beads
and is measured in units of $\varepsilon$. Here, we keep $\varepsilon_{\rm d} = \varepsilon$
and the Boltzmann constant, $k_B$ is taken as unity.  The~interaction between the droplet beads
and the substrate is realized through an LJ 9--3 potential, where the exponents result from the integration
of the LJ 12--6 potential over the substrate~\cite{Forte2014,Israelachvili,Theodorakis2015,Theodorakis2015b}. 
Hence, the~substrate is only implicitly present in the system representing a smooth and
unstructured substrate of `infinite' thickness. The~LJ 9--3 potential, $U^{\rm 9-3}(r)$, 
between each bead and the substrate reads
\begin{equation}\label{eq:LJpotential93}
U^{\rm 9-3}(z) = 4\varepsilon_{\rm s} \left[  \left(\frac{\sigma_{\rm  s}}{z}
\right)^{9} - \left(\frac{\sigma_{\rm s}}{z}  \right)^{3}    \right],
\end{equation}

\noindent
where the parameter $\varepsilon_{\rm s}$ is used to vary the interaction between the substrate
and the droplet beads, while $\sigma_{\rm s} = \sigma_{\rm d}$ {$=\sigma$ for simplicity}. 
The distance,
$z$, is simply the distance between the substrate and the droplet beads in the direction that 
is normal to the substrate ($z$ direction). As~in the case of the LJ 12--6 potential, the~LJ
{9--3} potential is cut and shifted at the same cutoff distance, 
{i.e.,} $z_{\rm c}=2.5\sigma$.

\begin{figure}[H]
\includegraphics[width=0.5\textwidth]{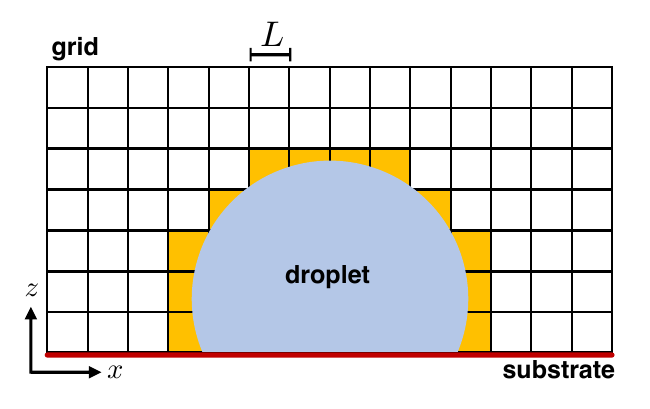}
\caption{\label{fig:1}
Schematic diagram of the system setup in two dimensions (2D)
for the sake of simplicity and clarity. The~grid has a size $L$. 
The cells at the liquid--vapour interface that are relevant for the
addition and the removal of particles are shaded with yellow color.
}
\end{figure}

The simulation approach of this study is designed to harvest the advantages of off-lattice MC
methods and the same time be suitable for investigating nucleation and evaporation phenomena at the molecular scale. The~bulk of the droplet is simulated by using the
standard NVT MC simulation method, but~the interface of the droplet is continuously tracked 
and treated differently. In~particular, a~three-dimensional (3D) grid is created, which is 
used to identify the liquid--vapour (LV) interface of the droplet by tracking the density of
the beads in the grid cells. A~similar concept has been used in the case of the volume of
fluid method~\cite{tryggvason_scardovelli_zaleski_2011}. However, in~our case this is only 
applied to identify the LV interface of the droplet (Figure~\ref{fig:1}). Then, beads belonging to the interface
are treated with an MC approach based on a Hamiltonian that involves the chemical potential, 
as, for~example, in~the case of the Rabani~et~al. model~\cite{Rabani2003}.
Our approach can be readily extended to incorporate other concepts, such as that of a varying
chemical potential depending on the droplet radius and time, as~in the case of the Zhang
{et al.} model~\cite{Zhang2016}. In~addition, the~model can be used with different
force-fields (including, also, all-atom force-fields), which renders
it particularly suitable for multicomponent systems ({e.g.,} liquid droplets with
nanoparticles). External fields ({e.g.,} electric field, gravity, {etc}.) 
can easily be added to the model { as additional energy terms}. While the method is described for MC simulations in 
this study, possible extensions based on the molecular dynamics (MD) method are 
conceivable~\cite{Tang2018,Tang2019a,Tang2019b}.

A cubic grid of {mesh} size $L > r_{\rm c}$ is initially created across the whole simulation domain
with periodic boundary conditions applied in all directions. In~the following, $L=4\sigma$.
Each cubic cell of the grid is
assigned beads according to their positions and the density of the cells is calculated. Cells
with a density below a certain threshold, $\rho_{\rm c}$ ({e.g.,} below the bulk density
of the droplet), will be identified as cells that contain particles belonging to the LV
interface. While the use of the grid facilitates the identification of the droplet interface,
it also provides an efficient way of finding the neighboring particles during the calculation
of the system energy by searching only the neighboring cells. In~our case, this is implemented
by using for each cell a Python dictionary that holds its neighboring cells.
The interaction energy of the system is described by the following Hamiltonian,
\begin{equation}\label{eq:Hamiltonian}
    H = \sum_{<\rm ij>} U^{\rm 12-6}_{\rm ij}(r) + \sum_{<\rm i>} U^{\rm 9-3}_{\rm i}(z)  -  \mu N,
\end{equation}

\noindent
where $<\rm ij >$ indicates interactions between all pairs of atoms that are found at
distances, $r$, smaller than the cutoff, $r_{\rm c}$. Similarly, the~sum over each bead
$<i>$ refers to the interaction of each bead with the substrate, when the distance between the
bead and the substrate is smaller than the cutoff, $z_{\rm c}$. $N$ is the number of particles in
the system and $\mu$ is the chemical potential, which is the energy cost when adding or removing
a particle to the system. This is a property of the particular component, 
which implies that the simulation
of multicomponent systems would require the definition of the chemical potential for each
particle type. In~practice, the~chemical potential here reflects the tendency for evaporation
or nucleation and is measured in units of $\varepsilon$. In~particular, more negative $\mu$
values would
favor evaporation, while increasing $\mu$ would favor nucleation.
The generation of subsequent states of the system is based on the realization of local MC 
moves and the addition or removal of beads to realize the evaporation and nucleation phenomena at the droplet interface. The~new state of the system is accepted by using the
Metropolis probability, \mbox{$p_{\rm acc} = \min [1,\exp(-\Delta H/K_BT)]$}, where $\Delta H$ is 
the energy difference between a new attempted state and the current state of the system. 
In the following, we discuss in detail how this framework is specifically implemented in
the nucleation and evaporation~cases.

\subsection{Nucleation}
After the initialization of the grid and the assignment of each particle
to the respective grid cell, the~system advances to subsequent states by randomly choosing a
particle from the droplet and realizing local MC particle moves. As~usual, the~number of such
attempts corresponds to the number of particles in the system. The~decision of accepting the
new state of the system is based on the Metropolis criterion. Then, an~attempt to add a new
particle to the system takes place. A~particle is randomly chosen and its cell and neighboring
cells are identified. An~attempt to add a new particle in these cells takes place. If~the new
particle is placed at a distance $r_{\rm n}$ ($\sigma < r_{\rm n} < r_{\rm c} $), then the 
new state is accepted according to the Metropolis criterion by considering the Hamiltonian of
\mbox{Equation~(\ref{eq:Hamiltonian})} and the MC cycle is completed. In~this approach, one can tune 
the distance threshold, $r_{\rm n}$, the~ratio between the attempts of local moves and adding
new particles, the~chemical potential, $\mu$, the~temperature of the system, $T$, and~the size
of the grid cells, $L$. The~detailed-balance condition would require an equal probability to
remove a particle from the system. However, one may consider in our case that this probability
is incorporated in the choice of the chemical potential, $\mu$, of~the system. Moreover, 
the chosen
simulation protocol would speed up the study of the droplet nucleation in this case, which is
itself a non-equilibrium process for the system. In~the following, the~protocol for 
evaporation can be used as well for studying nucleation phenomena by increasing the value of the chemical potential, in~this way favoring the addition of particles to the~system.

\subsection{Evaporation} 
The grid is initialized and particles are assigned to cells according to
their positions similar to what is done in the case of nucleation. Apart from the standard
local moves for all system beads, additional attempts to add and remove a particle take place
in each MC cycle for cells being at the LV interface. In~particular, a~bead is randomly
selected and its cell is identified. If~the density of the cell is below a density threshold,
$\rho_{\rm c}$, this cell contains beads of the LV interface. Then, a~standard local
move is attempted. If~the new position of the bead is within a threshold distance, $r_{\rm n}$,
($\sigma < r_{\rm n} < r_{\rm c} $ from its neighbours, the~new position of the bead is
accepted according to the Metropolis criterion. If~the distance between the particle and all
its neighbors is larger than $r_{\rm n}$, the~selected bead is removed according to the
Metropolis probability and considering the Hamiltonian of Equation~(\ref{eq:Hamiltonian}). In~the
latter case, particles that have moved far from the droplet according to the distance
criterion, $r_{\rm n}$, are considered as evaporated particles. From~the cell of the 
previously-selected particle, we randomly choose a bead and attempt to remove it. This 
attempt is accepted according to the Hamiltonian of Equation~(\ref{eq:Hamiltonian}) and the
Metropolis criterion. Finally, an~attempt to add a new bead in the selected cell takes place.
The addition of the new particle is also accepted according to the Hamiltonian of
Equation~(\ref{eq:Hamiltonian}) and the Metropolis~criterion. 

The above simulation protocols take advantage of the flexibility of the MC approach in adding
and removing particles from the system, as~well as the use of the grid cells. These protocols
constitute the basis upon which further methodology developments could be proposed in this area
in the future, including methods based on MD or even multiscale protocols~\cite{Smith2018}.
Therefore, various modifications of the algorithms are expected for improving and adjusting 
the model to the particular problem at hand. For~this reason and for better understanding the
behavior and limits of the method, we provide the implementation of our approach as a Python
code. In~the following, we present results from our simulations that illustrate the impact of
the various parameters on the~model.

\section{Results}

We have performed a broad exploration of the model parameters, $T$, $\varepsilon_{\rm s}$,
$r_{\rm n}$, $\rho_{\rm c}$, and~$\mu$. In~particular, we have considered the following 
range of values for each parameter: $0.2 \varepsilon/k_B \leq T \leq 1.0\varepsilon/k_B $, 
$1.1 \sigma \leq r_{\rm n} \leq 1.5\sigma$, 
$0.6 \sigma^{-3} \leq \rho_{\rm c} \leq 0.9 \sigma^{-3}$, 
and $-4.0 \varepsilon \leq \mu \leq 2.0 \varepsilon$.
Of course, the~choice of the system temperature, $T$, affects proportionally all the related
energy parameters of the model. Based on our analysis, we have found that the choice of $r_{\rm n}$
plays an important role in the model. The~influence of the latter and all parameters of the model will be discussed in more detail in the~following.

\begin{figure}[H]\widefigure
\includegraphics[width=0.9\textwidth]{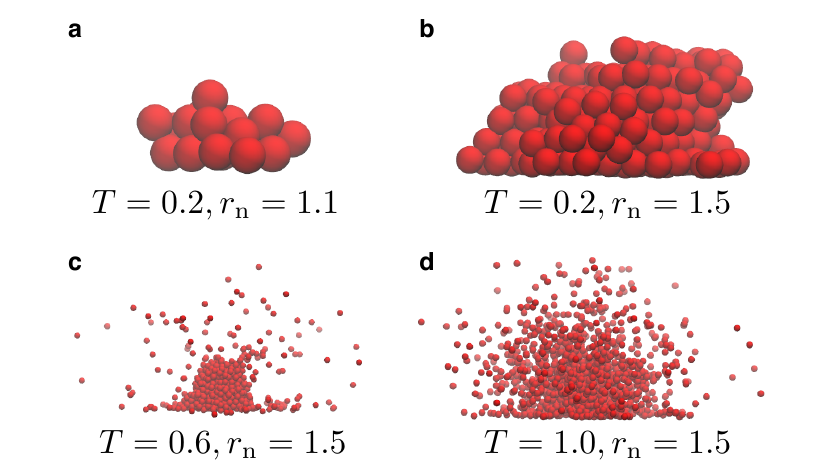}
\caption{\label{fig:2} (\textbf{a--d}) 
 Snapshots obtained from the nucleation
algorithm for various temperatures, $T$ (in units of $\varepsilon/k_B$) and $r_{\rm n}$
(in units of $\sigma$), as~indicated. For~all cases, $\mu = -1.0\varepsilon$, 
$\varepsilon_{\rm s}=1.5\varepsilon$,
and $\rho_{\rm c}=0.7\sigma^{-3}$. Snapshots are taken after the realisation of $10^4$ MCS. 
The scale of each snapshot has been
adjusted in order to make the snapshots visually clearer. }
\end{figure}

\begin{figure}[H]
\widefigure
\includegraphics[width=0.9\textwidth]{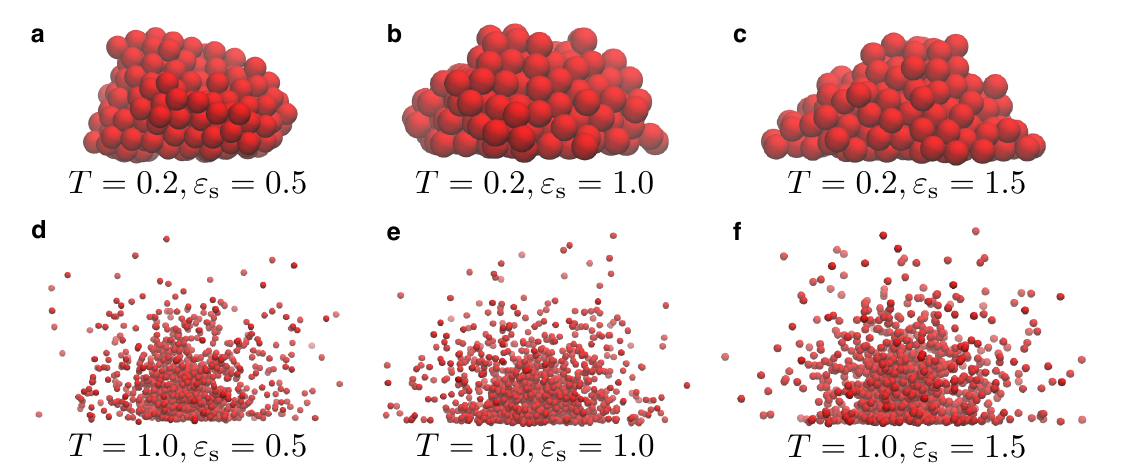}
\caption{\label{fig:3} (\textbf{a--f}) 
Snapshots obtained from the nucleation
algorithm for various temperatures (in units of $\varepsilon/k_B$) and
droplet--substrate attraction strength $\varepsilon_{\rm s}$ (in units of $\varepsilon$),
as indicated. For~all cases, $\mu = -1.0\varepsilon$, $r_{\rm n}=1.4\sigma$,
and $\rho_{\rm c}=0.7\sigma^{-3}$. The~scale of each snapshot has been
adjusted in order to make the snapshots visually clearer. }
\end{figure}

\begin{figure}[H]
\widefigure
\vspace{-4cm}
\includegraphics[width=1.0\textwidth]{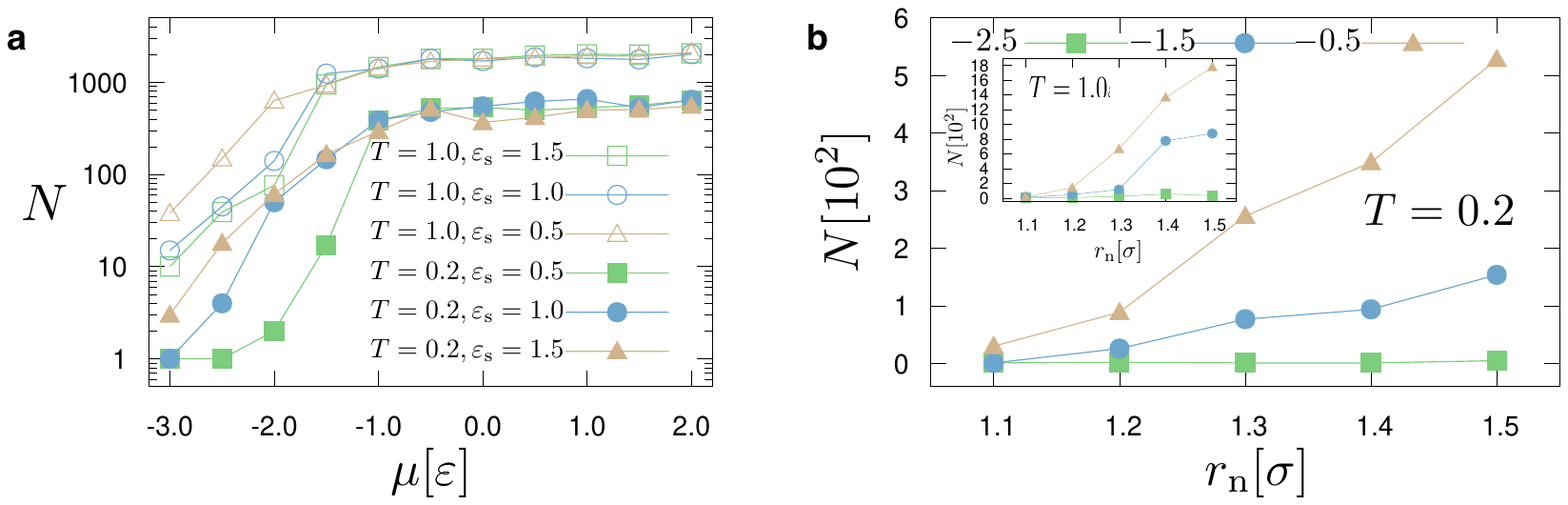}
\vspace{-4cm}
\caption{\label{fig:4}
(\textbf{a}) Dependence of the number of particles, $N$, on~the chemical
potential, $\mu$, for~different temperature, $T$ (in units of $\varepsilon/k_B$)
and attraction strength, $\varepsilon_{\rm s}$ (in units of $\varepsilon$), as~indicated. 
$r_{\rm n}=1.5\sigma$ and $\rho_{\rm c}=0.7\sigma^{-3}$;
(\textbf{b}) Dependence of the number of particles, $N$, in~the system as a function of the parameter
$r_{\rm n}$ for different cases of the chemical potential, $\mu=-0.5, -1.5$, and~$-2.5\varepsilon$, as~indicated. Main panel shows data for temperature, $T=0.2\varepsilon/k_B$, while inset
presents data for $T=1.0\varepsilon/k_B$.
For all cases, $\rho_{\rm c}=0.9\sigma^{-3}$, $\varepsilon_{\rm s}=1.5\varepsilon$.
}
\end{figure}

\begin{figure}[H]
\widefigure
\includegraphics[width=0.98\textwidth]{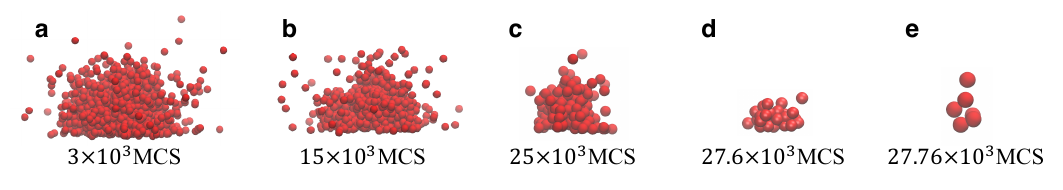}
\caption{\label{fig:5} (\textbf{a--e}) 
Various snapshots of droplet evaporation at different times (measured in MCS), as~indicated. 
Here, $T=1.0\varepsilon/k_B$, $\mu = -1.0\varepsilon$, $\varepsilon_{\rm s}=1.0\varepsilon$, 
$r_{\rm n}=1.5\sigma$, and~$\rho_{\rm c}=0.7\sigma^{-3}$. The~scale of each snapshot has been
adjusted in order to make the snapshots visually clearer.
}
\end{figure}

\begin{figure}[H]
\widefigure
\includegraphics[width=0.98\textwidth]{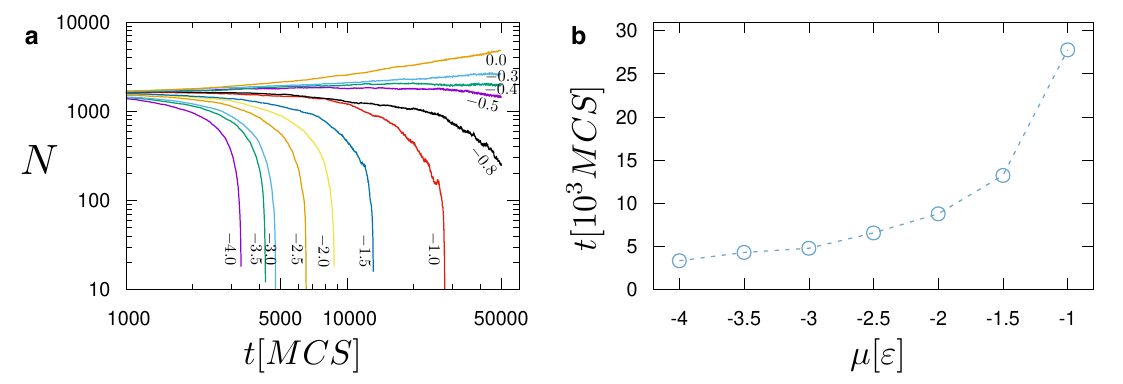}
\caption{\label{fig:6} (\textbf{a}) 
Dependence of the number of particles during evaporation for
different values of the chemical potential, $\mu$ ($-4.0\leq\mu\leq0.0$)
in units of $\varepsilon$, as~indicated,
as a function of time measured in the number of Monte-Carlo steps (MCS).
The liquid (droplet) and the vapour phases are in equilibrium for $\mu = -0.4 \varepsilon$; 
(\textbf{b}) Time (measured in MCS) required for the full evaporation of the droplet, as~a function of the
chemical potential, $\mu$. 
$T=1.0\varepsilon/k_B$, $r_{\rm n}=1.5\sigma$, $\rho_{\rm c}=0.7\sigma^{-3}$,
and $\varepsilon_{\rm s}=1.0\varepsilon$.
}
\end{figure}

\begin{figure}[H]
\widefigure
\includegraphics[width=.98\textwidth]{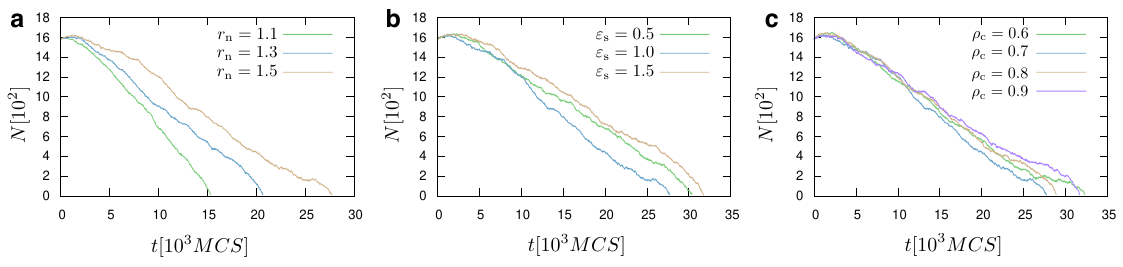}
\caption{\label{fig:7} Evaporation of a droplet at temperature $T=1.0\varepsilon/k_B$,
expressed through the number of beads,
$N$, of~the system as a function of time, $t$, for~different values of 
$r_{\rm n}$ (\textbf{a}), $\varepsilon_{\rm s}$ (\textbf{b}), and~$\rho_{\rm c}$ (\textbf{c}), as~indicated on 
each graph. $\mu=-1.0\varepsilon$, $\rho_{\rm c}=0.7\sigma^{-3}$, 
and $\varepsilon_{\rm s}=1.0\varepsilon$. $r_{\rm n}=1.5\sigma$ in panels (\textbf{b},\textbf{c}).
}
\end{figure}

\begin{figure}[H]
\widefigure
\includegraphics[width=0.98\textwidth]{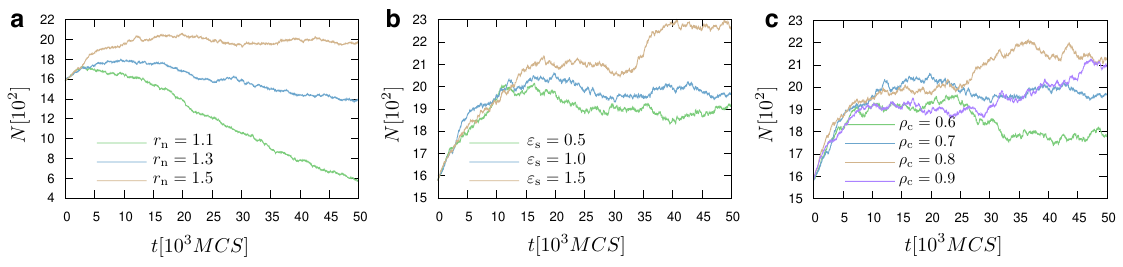}
\caption{\label{fig:8} Same as in Figure~\ref{fig:6}, but~$\mu=-0.4\varepsilon$. An~equilibrium between the liquid droplet and the surrounding vapour exists for
$r_{\rm n}=1.5\sigma$. $r_{\rm n}=1.5\sigma$ in panels (\textbf{b},\textbf{c}), while cases with different $r_{\rm n}$ are shown in panel (\textbf{a}), as~indicated. 
}
\end{figure}

\begin{figure}[H]
\widefigure
\includegraphics[width=0.98\textwidth]{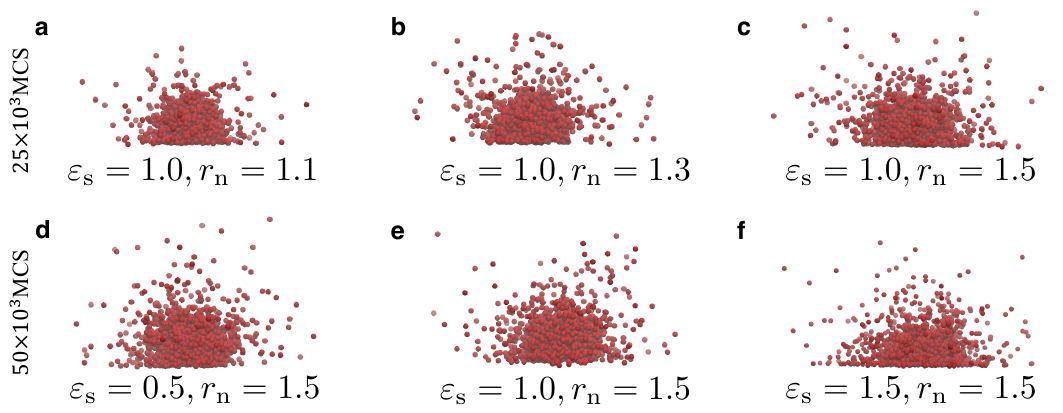}
\caption{\label{fig:9} 
\textbf{(a--f}) Snapshots of the system at $T=\varepsilon/k_B$ for different 
values of $\varepsilon_{\rm s}$
and $r_{\rm n}$ at different times (top panels: $25\times10^3$ MCS; bottom panels: 
$50\times10^3$ MCS). For~all cases, $\mu=-0.4\varepsilon$ and $\rho_{\rm c}=0.7\sigma^{-3}$.
The scale of each snapshot has been adjusted in order to make the snapshots visually clearer.
}
\end{figure}

Figure~\ref{fig:2} illustrates representative snapshots for the model for various choices for
the temperature and the parameter $r_{\rm n}$. We remind the reader that $r_{\rm n}$ is used 
to distinguish whether a particle at the LV interface belongs to the liquid droplet or is 
part of the vapor. This value is larger than the size of the droplet beads, 
$\sigma_{\rm d}$, and~up to $1.5 \sigma$ in our case, which is a distance within the 
first and second interaction shells of particles in the bulk. Our simulations start by
initially placing a single particle onto the substrate, whence the droplet starts to grow. 
As the droplet grows, we observe that vapor particles are absent at lower temperatures 
(e.g., $T=0.2 \varepsilon/k_B$), or~their presence is negligible during the simulation. 
Moreover, the~ addition of new particles to the droplet takes place faster when $r_{\rm n}$ is
larger. In~general, we have found that values of $r_{\rm n}$ larger than 1.1 are required to
start the nucleation process at larger temperatures, when the rest of the model parameters are 
kept the same. As~the temperature of the system increases, we also observe the presence of vapor
around the liquid droplet. Hence, in~this sense the model works as expected by only preserving
the vapor close to the LV interface. Moreover, we can clearly distinguish the boundaries of the 
LV interface. As~the temperature further increases, thermal fluctuations become more pronounced
in the system both at the bulk and the LV interface.  In~all cases, vapor exists only close to 
the LV interface and particles far away from the interface will eventually be removed 
during the simulation. As~a result, the~computational time of the simulation spent on vapor 
particles in the case of our model is rather small. In~the snapshots of Figure~\ref{fig:2}, we can 
also see the impact of the substrate potential, which is rather large in these particular
cases, namely $\varepsilon_{\rm s}=1.5\varepsilon$. More specifically, we can observe the
distortion of the droplet contact-line at higher temperatures with beads lying nearby 
onto the substrate.
The influence of the substrate can be visually summarized by the snapshots of Figure~\ref{fig:3}
for the lowest ($T=0.2\varepsilon/k_B$) and the highest ($T=1.0\varepsilon/k_B$) temperatures
considered in our study. We found that the strength of the substrate potential has
a small effect on the growth rate of the droplet, independently of the temperature. 
However, the~final configurations will be different under the influence of the
substrate potential, especially close to the contact line. For~example, we can observe the
formation of a precursor layer at the contact line of the droplet at higher temperatures.
Moreover, smaller contact lines are observed for the droplet for both temperatures, when
the strength of the substrate potential is~larger.
\end{paracol}
\nointerlineskip
\begin{paracol}{2}
\switchcolumn

Figure~\ref{fig:4} presents results for the dependence of the number of particles in the system as
a function of the chemical potential, $\mu$, and~the distance parameter, $r_{\rm n}$. As~mentioned before, these parameters significantly affect the behavior of the system.
We can observe that the number of particles increases as a function of the chemical potential
during a simulation of $10^4$ MCS. Above~a certain value, namely $\mu=-1.0\varepsilon$, the~influence of the chemical potential is small, independently of the temperature of the system.
Moreover, we observe that the number of particles, $N$, is rather larger in the case of a smaller
substrate--droplet interaction, that is $\varepsilon_{\rm s}=0.5\varepsilon$. A~similar behavior
is observed in the case of a higher temperature ($T=1.0\varepsilon/k_B$). The~dependence of the
number of particles of the system, $N$, on~$r_{\rm n}$ is also significant. Within~the time of
the simulation ($10^4$ MCS), we can see that  small values of $r_{\rm n}$ restrict the addition of
new particles to the droplet. As~$r_{\rm n}$ becomes larger, we observe a greater ability of 
adding particles to the system. This ability also depends on the value of the chemical potential.
In particular, the~higher the chemical potential, the~higher is the dependence on $r_{\rm n}$. 
By examining the structures of all our cases, we have observed that the value of $r_{\rm n}$
affects the rate of droplet growth, but~it generally does not { influence} the droplet configurations, 
for $r_{\rm n}>1.1${ $\sigma$}. We have also seen that $r_{\rm n}=1.1${ $\sigma$} rather hinders the growth of the
droplet on the substrate and the formation of droplets has not been possible for various choices 
of model parameters within the available time of the simulation when the temperature increases.
The above discussion is consistent throughout the extensive parameter exploration considered in
this~study.
\end{paracol}
\nointerlineskip
\begin{paracol}{2}
\switchcolumn

\end{paracol}
\nointerlineskip
\begin{paracol}{2}
\switchcolumn

We now turn our discussion to the evaporation model, which is the main focus of our work. 
Figure~\ref{fig:5} presents various snapshots during the evaporation process of a droplet
for a particular case. The~initial configuration of the system is a droplet that contained 1578~particles and was created with our nucleation algorithm. At~each stage of the evaporation process, we can clearly distinguish the bulk of the droplet
and the surrounding vapor in the system. During~evaporation, the~droplet changes configurations
by using local MC moves and exchanging particles at the LV with the surrounding vapor phase. 
The algorithm produces consistent results  independently of the droplet size and until the 
droplet has fully~evaporated. 
\clearpage
\end{paracol}
\nointerlineskip
\begin{paracol}{2}
\switchcolumn

The rate of evaporation depends on the choice of the chemical potential. As~shown in
Figure~\ref{fig:6}, more negative values of the chemical potential lead to faster evaporation.
In contrast, values larger than $\mu=-0.4\varepsilon$ will lead to the addition of particles to 
the droplet. For~the particular choice of parameters, we observe an equilibrium between the 
liquid and the surrounding vapor particles for $\mu=-0.4\varepsilon$. Establishing such
an equilibrium is a key element for the success of our model. This indicates that the
liquid droplet can be reliably simulated while coexisting with the surrounding vapor particles.
In all evaporation cases
($\mu<-0.4\varepsilon$), we have found that the droplet initially evaporates at a slower pace. 
When the
droplet size reaches about 250 particles in this case, then its evaporation accelerates (\mbox{Figure~\ref{fig:6}}).
Hence, we can distinguish two different behaviors, which are determined by the size of the 
droplet. By~examining the snapshots of the system at each evaporation stage ({e.g.,}~see
Figure~\ref{fig:5}), we have found that the pace of evaporation is only affected by the chemical
potential for a given set of model parameters. The~model produces consistent results and 
configuration changes take place as expected. As~the chemical potential increases, droplet
evaporation requires larger times, which grow exponentially as the chemical potential reaches 
the point that the liquid and vapor particles are in dynamic equilibrium (Figure~\ref{fig:6}b). 
\end{paracol}
\nointerlineskip
\begin{paracol}{2}
\switchcolumn

Our evaporation model is not sensitive to the choice of the parameter $\rho_{\rm c}$
(Figure~\ref{fig:7}), as~it has also been found in the case of the nucleation protocol. In~addition, the~influence of the wall potential has a small effect when $\varepsilon_{\rm s}=0.5$ 
and $1.0\varepsilon$, while the case $\varepsilon_{\rm s}=1.5\varepsilon$ would lead to a slight delay of the complete
droplet evaporation, due to the extra energy that the substrate provides to the particles. 
However, $r_{\rm n}$ would significantly affect the evaporation process. In~particular, smaller
values of $r_{\rm n}$ ({e.g.,} $r_{\rm n}=1.1\sigma$) would lead to a faster evaporation of the
droplet, since beads at a distance $r_{\rm n}=1.1\sigma$ away from the LV will be already considered 
as part of the vapour. In~contrast, larger values of $r_{\rm n}$ ({e.g.,} $r_{\rm n}=1.5\sigma$,
which is also a more natural choice as it includes the first shell of neighbours), would lead to
larger times for the complete evaporation of the droplet. Hence, as~in the case of  nucleation, 
the choice of $r_{\rm n}$ is crucial in the case of the evaporation~algorithm.
\end{paracol}
\nointerlineskip
\begin{paracol}{2}
\switchcolumn

The choice of $r_{\rm n}$ can affect the dynamic equilibrium of the system. Hence, once $r_{\rm n}$
is chosen, it has to remain the same throughout a study. For~example, when $\mu=-0.4\varepsilon$
and $r_{\rm n}=1.5\sigma$, we observe the dynamic equilibrium between the liquid and the vapor phases
(Figure~\ref{fig:8}). However, when another value of $r_{\rm n}$ 
({e.g.,} $r_{\rm n}=1.2\sigma$) is chosen, the~value of the chemical potential required to
establish the equilibrium between the droplet and its surrounding vapor particles, would also
change. Figure~\ref{fig:9} illustrates characteristic snapshots at time $25\times10^3$ MCS for
different $r_{\rm n}$ cases. As~previously mentioned, the~evaporation process is faster in the case of smaller $r_{\rm n}$, which would lead to a smaller size of the droplets. Comparing the configurations during evaporation for the different cases of $r_{\rm n}$, we do not
see any significant structural differences. Hence, the~dynamic equilibrium between the liquid 
and the vapor phases can be obtained again by a proper choice of the chemical potential, $\mu$.
For the case that a dynamic equilibrium is established for $r_{\rm n}=1.5\sigma$, we have included a
movie as Supplementary Information. We have also found that changes in $\rho_{\rm c}$ or
$\varepsilon_{\rm s}$ would still maintain the equilibrium in the system for the same value of 
the chemical potential (Figure~\ref{fig:8}). Moreover, different choices for $\varepsilon_{\rm s}$ may slightly shift
this equilibrium to larger or smaller droplets without the requirement to change the value of the
chemical potential. In~particular, we observe that larger values of $\varepsilon_{\rm s}$ would
favour a larger droplet size (Figure~\ref{fig:8}). The~effect of the substrate attraction 
strength on the droplet shape for droplets in {dynamic} equilibrium between the vapour and the liquid
phases, in~particular at the contact line, can be seen in the snapshots of Figure~\ref{fig:9}.
\end{paracol}
\nointerlineskip
\begin{paracol}{2}
\switchcolumn

\unskip 

\clearpage
\end{paracol}
\nointerlineskip
\begin{paracol}{2}
\switchcolumn
\vspace{-12pt}

\section{Conclusions}
In this study, we have proposed an off-lattice approach, which can be used to simulate nucleation
and evaporation phenomena of droplets at the molecular scale. The~model can be easily extended 
for any forcefield, be it atomistic or coarse-grained. We have taken advantage of the flexibility
of the MC approach combined with a 3D grid. The~grid is used to identify the position of the LV
interface, where the addition and removal of particles take place. Moreover,
vapor particles far from the LV interface are naturally removed by the algorithm during the
simulation, which {also makes} our approach computationally efficient. The~model works
as we expected with the chemical potential controlling the processes and nicely capturing the liquid
droplet as well as the vapor particles around the droplet. Moreover, our evaporation protocol 
is able to establish a dynamic equilibrium between the droplet and the surrounding vapor
particles. Hence, evaporation, a dynamic equilibrium between liquid and vapor, as~well as nucleation
phenomena can be modeled based on this approach. By~examining a broad range of values for 
the model parameters, we have found that the
parameter $r_{\rm n}$ should be chosen carefully and remain constant during the study of a
particular problem at hand. Further improvements of the model are conceivable given the 
flexibility of the MC approach. To~facilitate this, we have provided a Python implementation of 
the model. { Such model extensions have to be adjusted to the particular
applications and the choice of the forcefield. Possible applications include the
study of evaporation phenomena in complex systems, for~example, liquid droplets with 
any kind of molecules, nanoparticles, etc. under different conditions
({e.g.,} external fields). By~using the framework of this study, these systems
can be modeled by using any force-field ({e.g.,} all-atom~\cite{Berendsen1987} and 
coarse-grained~\cite{Souza2021,Lafitte2013} models) at the particular temperature that the 
forcefield was obtained. Hence, our approach provides new areas
of application for popular force-fields in nucleation and evaporation phenomena without the need
for changing their parameters, since the processes are controlled at the LV interface 
by the chemical potential. The~approach is flexible and can also be extended for more complex
system setups suitable for studying bubbles or heat transfer processes between the
substrate and the liquid and the vapor. In~all these cases, different Monte-Carlo 
schemes can be used for generating the configurations and guaranteeing that certain
criteria are met. We expect that in these cases the grid approach will be able to 
track the boundary between the liquid and the vapor phases, 
which is the crucial element for the success of our approach. } 
Thus, we anticipate that our method will open new opportunities for the molecular-level simulation of nucleation and evaporation~phenomena.

\vspace{6pt} 

\supplementary{The following are available online at \linksupplementary{s1}, Implementation of the method in Python3.8 Programming language including an example of an input file with associated documentation and a script to visualise the trajectories in povray. A~movie demonstrating the dynamic equilibrium between the liquid droplet and the surrounding vapour. In~this case, $T=\varepsilon/k_B$, $\mu=-0.4\varepsilon$, $\varepsilon_{\rm s}=1.0\varepsilon$, $r_{\rm n}=1.5\sigma$, and~$\rho_{\rm c}=0.7\sigma^{-3}$.}


 \authorcontributions{
Conceptualization, P.E.T. and B.L.; methodology, P.E.T.; software, P.E.T.; validation, P.E.T.; investigation, P.E.T., Y.W., A.C. and B.L.; resources, P.E.T.; data curation, P.E.T.; Analysis, P.E.T., Y.W., A.C. and B.L.; writing--original draft preparation, P.E.T.; writing--review and editing, P.E.T., Y.W., A.C. and B.L.; visualization, P.E.T.; supervision, P.E.T. and B.L.; funding acquisition, P.E.T. and B.L.}

\funding{
This project has received funding from the European Union’s Horizon 2020 research and innovation programme under the Marie Skłodowska-Curie grant agreement No. 778104. This research was supported in part by PLGrid~Infrastructure.}

\institutionalreview{
N/A}

\informedconsent{
N/A}

\dataavailability{
The implementation of the method as a Python3.8 program is provided in the Supplementary Materials.}

\acknowledgments{This project has received funding from the European Union’s Horizon 2020 research and innovation programme under the Marie Skłodowska-Curie grant agreement No. 778104. This research was supported in part by PLGrid~Infrastructure.}

\conflictsofinterest{The authors declare no conflict of~interest.} 





\end{paracol}
\reftitle{References}




\end{document}